\documentclass[a4paper]{article}

\usepackage{INTERSPEECH2022}
\usepackage{amsmath}
\usepackage{multirow}

\DeclareMathOperator*{\argmax}{arg\,max}

\ninept
\title{Adversarial Reweighting for Speaker Verification Fairness}
\name{Minho Jin$^1$, Chelsea J.-T. Ju$^2$, Zeya Chen$^2$, Yi-Chieh Liu$^2$, Jasha Droppo$^2$, and Andreas Stolcke$^2$}
\address{
$^1$Amazon Web Services, Palo Alto, CA, USA\\
$^2$Amazon Alexa AI, Sunnyvale, CA, USA
}
\email{\{minhoj,juitij,zeyachen,ayliua,drojasha,stolcke\}@amazon.com}

\begin{document}

\maketitle
\begin{abstract}
We address performance fairness for speaker verification using the adversarial reweighting (ARW) method. 
ARW is reformulated for speaker verification with metric learning, and shown to improve results across different subgroups of gender and nationality, without requiring annotation of subgroups in the training data.
An adversarial network learns a weight for each training sample in the batch so that the main learner is forced to focus on poorly performing instances. Using a min-max optimization algorithm, this  method improves overall speaker verification fairness.
We present three different ARW formulations: accumulated pairwise similarity, pseudo-labeling, and pairwise weighting, and measure their performance in terms of equal error rate (EER) on the VoxCeleb corpus. 
Results show that the pairwise weighting method can achieve 1.08\% overall EER, 1.25\% for male and 0.67\% for female speakers, with relative EER reductions of 7.7\%, 10.1\% and 3.0\%, respectively. 
For nationality subgroups, the proposed algorithm showed 1.04\% EER for US speakers, 0.76\% for UK speakers, and 1.22\% for all others. 
The absolute EER gap between gender groups was reduced from 0.70\% to 0.58\%, while the standard deviation over nationality groups decreased from 0.21 to 0.19.
\end{abstract}

\section{Introduction}
Speaker verification means to determine whether an input speech utterance matches an enrolled speaker  \cite{reynolds2002overview, hansen2015speaker}. 
This is a key machine learning (ML) problem for voice assistants, as it enables the system to provide a personalized user experience. 
Given the broad acceptance of voice assistants (such as Amazon Alexa, Google Home, and Apple Siri) in everyday life, we consider it important that systems perform uniformly well for all user groups.  Problems with performance fairness for speech processing systems have been pointed out with respect  
to race \cite{koenecke2020racial} and nonnative accents \cite{wu2020see}.
Here we address performance fairness specifically for speaker recognition, and specifically, speaker verification, where fairness has not received much study so far.

One approach to avoid performance disparities for groups of speakers is to sample more data from underrepresented groups \cite{caton2020fairness, mehrabi2021survey, kamiran2010classification, bej2021loras, gupta2018proxy}.
Sampling from underrepresented groups is based on assumptions about their distribution. 
If those assumptions are incorrect, this can create its own biases relative to attributes such as gender or nationality.
Also, obtaining the required annotations may be difficult due to privacy concerns \cite{caton2020fairness}.
To address the fairness problem, many researches have been focusing on identifying the biases in the training data \cite{caton2020fairness, mehrabi2021survey}.
These include applications to face recognition \cite{samadi2018price, terhorst2021comprehensive}, image classification \cite{hwang2020exploiting}, speech recognition \cite{liu2021model}, natural language processing \cite{swinger2019biases, thompson2021bias}, and classification \cite{menon2018cost,oneto2019taking}.
For speaker verification, Fenu {\it et al.}~\cite{fenu2021fair} showed how data balancing in training can improve fairness. 
Toussaint and Ding \cite{toussaint2021sveva} demonstrated that an end-to-end ResNet speaker verification model suffers from performance divergence for different nationalities in VoxCeleb data.
Shen {\it et al.}~\cite{shenimproving} recently showed that using fusion networks trained on subgroups (e.g., male and female speakers) can improve speaker verification fairness, including for unbalanced training data. 

In this work, we focus on approaches that mitigate uneven performance for different data subsets without using additional attributes (such as gender and nationality) in the data, motivated by Laoti {\it et al.}~\cite{lahoti2020fairness}. 
Similarly to the distributionally robust optimization (DRO) method proposed by Hashimoto {\it et al.}~\cite{hashimoto2018fairness}, this approach aims to automatically identify the underperforming groups using an adversarial network, and adjusts their contribution to the training loss.
This was shown to not only reduce the performance disparity for minority groups, but to also improve performance overall. 
The adversarial network is trained so that it can identify and boost the weight of underperforming data using a min-max optimization \cite{lahoti2020fairness}:
\begin{equation}
J(\theta, \phi) = \min_{\theta} \max_{\phi} \sum_{(\mathbf{x}_i,y_i) \in D\times L} \lambda_{\phi}(\mathbf{x}_i, y_i) l(h_{\theta}(\mathbf{x}_i), y_i) \label{eq:J} \ ,
\end{equation}
where $D$ and $L$ are data and label sets, respectively. 
Given a data sample $\mathbf{x}_i$ with its label $y_i$, the loss is computed as  $l(h_{\theta}(\mathbf{x}_i), y_i)$ using the learner network parameters $\theta$.
The adversarial weight $\lambda_{\phi}(\mathbf{x}_i, y_i)$ is generated by the adversarial network with parameters $\phi$.
Given a learner network parameter $\theta$, the adversary parameters $\phi$  are first updated to maximize the weighted loss $J(\theta, \phi)$, which is expected to increase $\lambda_{\phi}(\mathbf{x}_i, y_i)$ for underperforming groups.
Then, $J(\theta, \phi)$ is minimized with respect to $\theta$ to reduce the training loss.

We address how to apply this min-max optimization for speaker verification with metric learning, more specifically with angular prototypical loss \cite{snell2017prototypical, Chung_2018}.
Unlike the classification problem in (\ref{eq:J}), where $l(h_{\theta}(\mathbf{x}_i), y_i)$ is determined by the data instance and its label, the loss for metric learning is defined using a batch of utterances from multiple speakers, and the ability of the learner to classify pairs of utterances.
For this purpose we propose three different formulations: accumulated pairwise similarity (APS), pseudo-labeling (PL) with K-means, and pairwise weighting (PW).
In APS, the adversarial network computes each speaker's weight using the aggregated similarity to other speakers. 
In PL, each speaker is labeled in an nsupervised manner using K-means, and the adversarial weights are a function of these pseudo-labels.
Finally, in PW, we make the adversarial weights a function of pairs of speakers.
We evaluate the various approaches in terms of overall and group-wise equal error rate (EER) for speaker verification on the VoxCeleb corpus.

Section \ref{sec:algorithm} describes the baseline speaker verification system and the proposed ARW algorithms for speaker verification. Section \ref{sec:experiment} describes our experiments, and Section \ref{sec:conclusion} summarizes the findings.

\section{Algorithms\label{sec:algorithm}}
\begin{figure}[t!]
  \centering
  \includegraphics[width=0.8\linewidth]{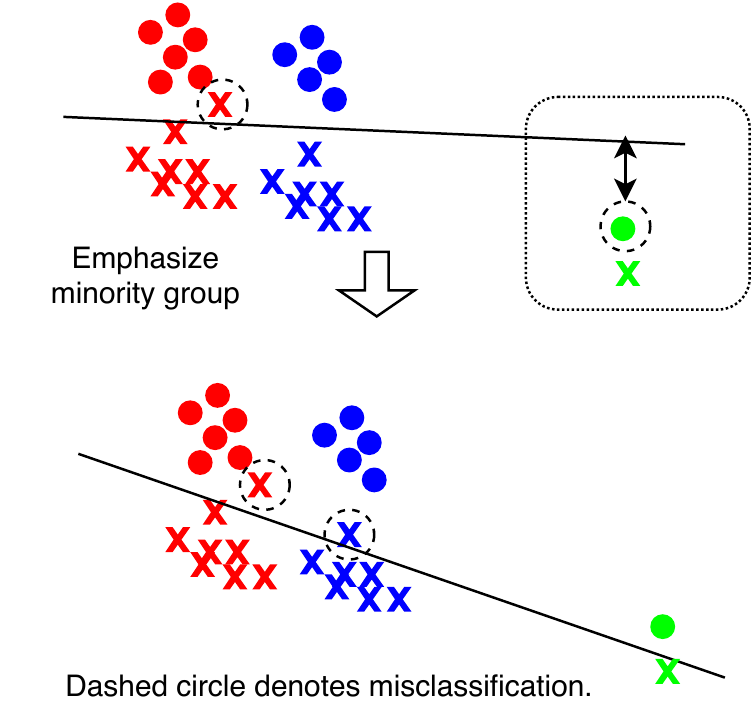}
  \caption{Representation bias in the training data, where each group is colored differently. For binary classification of two different classes of circle and cross, the bottom has better fairness than the top by considering the minority group, green.}
  \label{fig:representation_bias}
\end{figure}

\begin{figure}[t]
  \centering
  \includegraphics[width=0.7\linewidth]{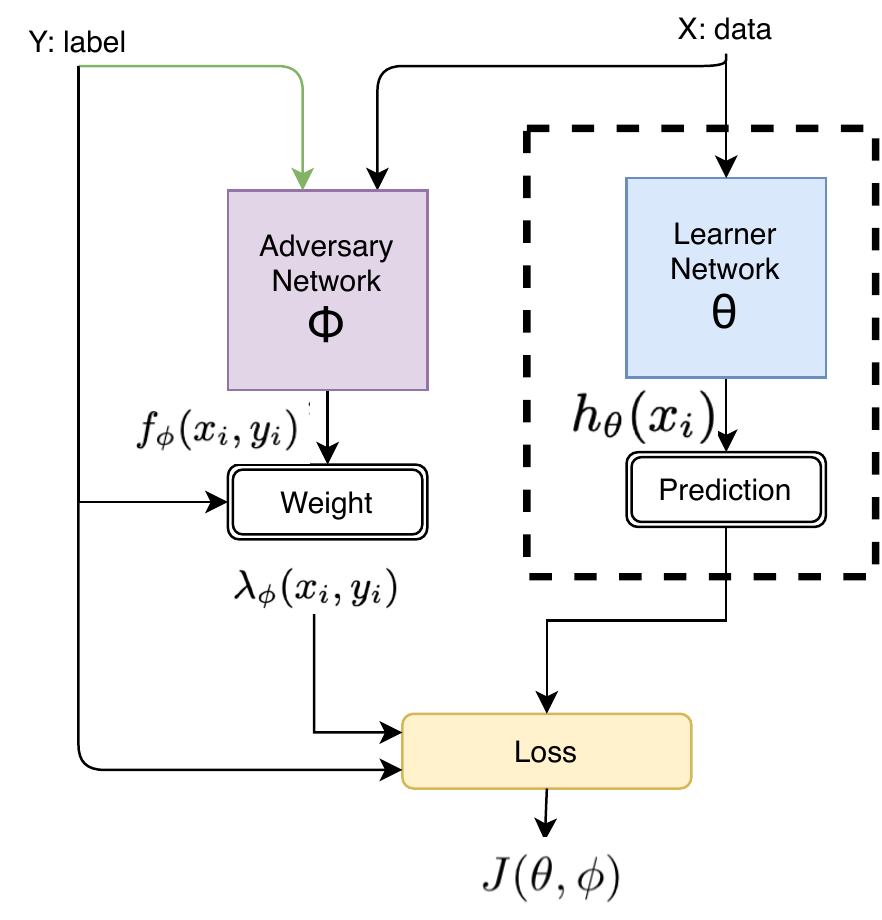}
  \vspace{-1em}
  \caption{ARW for classification. For inference, only the learner inside the dashed box is used.}
  \label{fig:arw}
\end{figure}

\begin{figure}[t]
  \centering
  \includegraphics[width=0.7\linewidth]{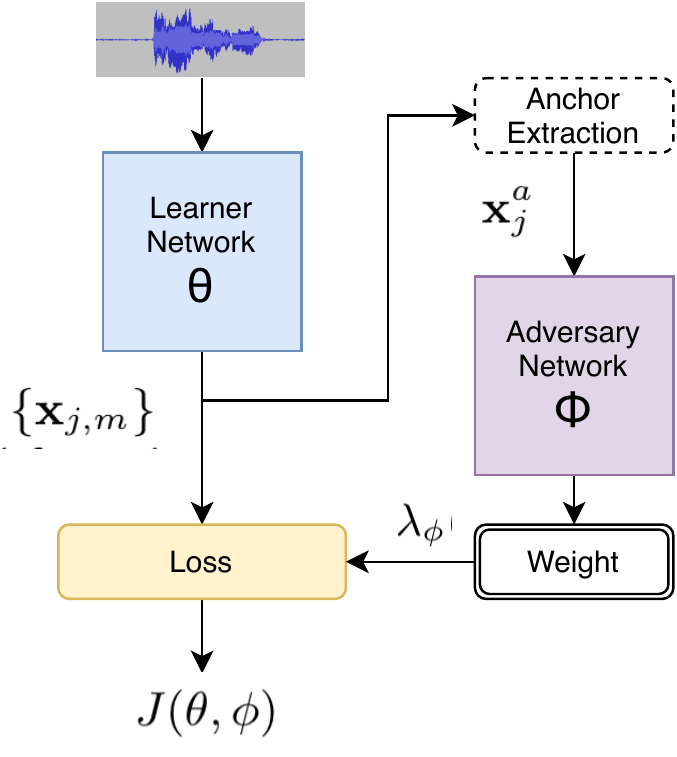}
  \vspace{-1em}
  \caption{ARW for speaker verification. We define the weight $\lambda_{\phi}$ differently for APS, PL, and PW.}
  \label{fig:arw_sv}
\end{figure}

\subsection{Speaker verification\label{sec:speaker_verification}}
The baseline speaker verification in our experiments uses utterance embeddings based on a ResNet model \cite{Chung_2018, chung20_odyssey, chung2020defence}. 
Given an audio input, the system extracts an embedding by feeding the Mel-band spectrogram of the audio to the ResNet model.
Given a pair of embeddings, it computes their cosine values to make a decision on whether they come from the same speaker.

When training the speaker verification model, we used angular prototypical (AP) and label loss \cite{snell2017prototypical, Chung_2018}. 
An input batch for training consists of $M$ utterances from $N$ speakers. 
For AP loss, we compute the anchor $\mathbf{x}^{a}_j$ and the query $\mathbf{x}^q_j$ for speaker $j$:
\begin{eqnarray}
\mathbf{x}^{a}_j &=& \frac{1}{M - 1} \sum_{m=0}^{M-2} \mathbf{x}_{j,m} \ ,\\
\mathbf{x}^{q}_j &=& \mathbf{x}_{j, M-1} \ ,
\end{eqnarray}
where $\mathbf{x}_{j,m}$ for $m \in \{0, \ldots, M-1 \}$ denotes the embedding extracted from the $m$th utterance from the speaker $j \in \{ 0, \ldots, N-1 \}$.
Using the anchor and the query, the similarity between a pair of speakers $(j, k)$ is computed as follows:
\begin{equation}
    \mathbf{S}_{j,k} = w \cos(\mathbf{x}^a_j, \mathbf{x}^{q}_k ) + b \ ,
\end{equation}
where
\begin{equation}
    \cos(\mathbf{x}^a_j, \mathbf{x}^{q}_k)= \frac{\mathbf{x}^{a}_j \cdot \mathbf{x}^{q}_k }{||\mathbf{x}^a_j||\  ||\mathbf{x}^{q}_k||} \,
\end{equation}
and where $w$ and $b$ are trainable weight and bias coefficients, respectively.
Finally the AP loss is computed according to
\begin{equation}
    L_p = \frac{1}{N} \sum_{j=0}^{N-1} L_{p,j} \ , \label{eq:L_p}
\end{equation}
where
\begin{equation}
    L_{p,j} = -\mathrm{log} \frac{e^{\mathbf{S}_{j,j}}} {\sum_{k=0}^{N-1} e^{\mathbf{S}_{j,k}}} \ .
    \label{eq:L_pj}
\end{equation}

For the label loss, all utterances in the training set are labeled by their speaker index in the training data. 
The embeddings are fed into a single-layer network whose output size is the same as the number of training speakers, and the loss is defined as the softmax loss given the true speaker index.
For this paper, we have adopted the best-performing model from \cite{heo2020clova}, which uses a combination of AP and label loss.

\subsection{Adversarial re-weighting algorithm\label{sec:arw}}
Fig. \ref{fig:representation_bias} illustrates how the representation bias of the training data affects fairness when classifying data into two classes, marked as circles and crosses. 
There are groups of red, blue and green data points, where green represents the minority group. 
The top of Fig.~\ref{fig:representation_bias} is an example of the optimal classification boundary for the entire training set.
However, the minority group (green) has 100\% classification error. 
If we can identify this green group and increase their contribution to the overall training loss, it may be possible to obtain a classification boundary as shown at the bottom of the figure. 
The two classification boundaries will give the same number of classification errors, marked as dotted circles, but the decision boundary at the bottom provides better fairness at the group level. 

In ARW, the adversarial network is trained along with the learner network so that it can learn to implicitly identify underperforming groups using the adversarial network. Fig.~\ref{fig:arw} shows the elements of the ARW algorithm for a classification task. 
Given input data $\mathbf{x}_i$ and label $y_i$, the learner makes a prediction of the class $y_i$. 
In \cite{lahoti2020fairness}, the weight $f_{\phi}(\mathbf{x}_i, y_i)$ is computed using a sigmoid for each class of $Y$ so that it ranges over $[0, 1]$, where $\phi$ is the set of parameters of the adversary network.
Finally, the weight $\lambda_{\phi}(\mathbf{x}_i, y_i)$ is normalized:
\begin{equation}
    \lambda_{\phi} (\mathbf{x}_i, y_i) = 1 + \frac{f_\phi(\mathbf{x}_i, y_i)}{\frac{1}{B} \sum_{j=0}^{B-1} f_\phi(\mathbf{x}_j,y_j)} \ ,
\end{equation}
where $B$ is the batch size. 
The training is done as in (\ref{eq:J}), where the adversary network $\phi$ tries to maximize the loss (i.e., upweighting the worst inputs or groups) and the learner (parameterized by $\theta$) tries to minimize the loss. 
In the implementation, we first train the learner for a sufficient number of epochs, and then iterate  maximization of $J(\theta, \phi)$ with respect to $\phi$ and minimization of $J(\theta, \phi)$ with respect to $\theta$.
In this original formulation, the adversary model output $f_\phi(\mathbf{x}_i, y_i)$ is defined for the case of single-input classification, and thus needs to be adapted for metric learning, as in speaker verification.

Fig.~\ref{fig:arw_sv} illustrates the reformulated ARW framework for speaker verification. Note that we feed the output embedding coming from the learner, rather than its input audio, to the adversary network. This simplifies the adversary network, since it takes in a fixed-length vector rather than a variable-length sequence of vectors. Within this general new framework, approaches differ in how the adversarial weights are utilized in the loss computation, as described next. 

\subsubsection{Accumulated pairwise similarity (APS)\label{sec:accumulate_pairwise}}

\begin{figure}[t]
  \centering
  \includegraphics[width=0.7\linewidth]{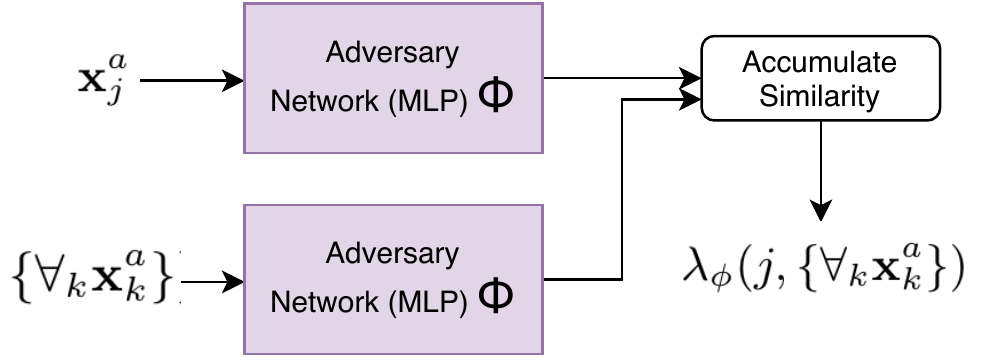}
  \vspace{-.5em}
  \caption{Computing ARW weights $\lambda_{\phi}(j, \{\forall_{k}\mathbf{x}_k^{a}\})$ for APS.}
  \label{fig:aps}
\end{figure}

Fig. \ref{fig:aps} illustrates how we compute weights in the APS formulation.
Unlike in the original ARW, the adversary network $f_{\phi}$ computes a vector instead of a scalar, mapping the anchor vectors to a new space.  Subsequently, the adversarial weights are computed from a comparison of the anchor to all speakers, using the inner product:
\begin{equation}
J(\theta, \phi) = \min_{\theta} \max_{\phi}  \sum_{\mathbf{x}_{j,k} \in D} \left(
  \sum_{j=0}^{N-1}  \lambda_{\phi}(j, \{\forall_{k}\mathbf{x}_k^{a}\}) L_{p,j} \right) \ . \label{eq:accumulated}
\end{equation}
We formulate two versions of APS adversarial weights; one based on inner product similarity:
\begin{equation}
    \lambda_{\phi}(j, \{\forall_{k}\mathbf{x}_k^{a}\} ) = 1+ \frac{\sum_{k} f_\phi(\mathbf{x}_j^{a}) \cdot f_\phi(\mathbf{x}_k^{a})} {\frac{1}{N}\sum_j \sum_{k} f_\phi(\mathbf{x}_j^{a}) \cdot f_\phi(\mathbf{x}_k^{a}) }\ , \label{eq:aps_bilinear} \\
\end{equation}
and one based on cosine similarity:
\begin{equation}
    \lambda_{\phi}(j, \{\forall_{k}\mathbf{x}_k^{a}\} ) = 1+ \frac{\sum_{k} e^{\cos(f_\phi(\mathbf{x}_j^{a}), f_\phi(\mathbf{x}_k^{a}))}} {\frac{1}{N}\sum_j \sum_{k} e^{\cos(f_\phi(\mathbf{x}_j^{a}), f_\phi(\mathbf{x}_k^{a}))} }\ . \label{eq:aps_exp}
\end{equation}
Here $f_\phi(\mathbf{x}_j^{a}) \in \mathbb{R}^H$ is the adversarial network, and $H$ is its output embedding dimension, a hyperparameter. 
We use the exponential of cosine similarity in (\ref{eq:aps_exp}) with an assumption that it is normally distributed.

\subsubsection{Pseudo-labeling with K-means (PL)\label{sec:pseudo labelling}}

\begin{figure}[t]
  \centering
  \includegraphics[width=0.7\linewidth]{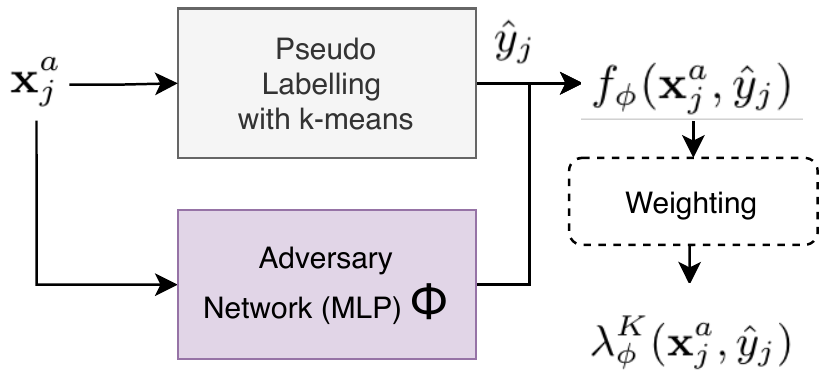}
  \vspace{-.5em}
  \caption{Computing weights $\lambda^K_{\phi}(\mathbf{x}_j^{a})$ for PL.}
  \label{fig:pl}
\end{figure}

Fig. \ref{fig:pl} illustrates how we compute weights with PL.
In this approach, we use the K-means algorithm \cite{macqueen1967some} to group speakers based on their embeddings, and label each speaker as one of $K$ clusters:
\begin{equation}
    \hat{y}_j = \argmax_{k \in \{0 , \ldots, K-1\}} \cos(\mathbf{c}_k, \mathbf{x}^a_j) \ ,\label{eq:hat_y_j}
\end{equation}
where $\mathbf{c}_k$ is the $k$th centroid. 
The learner is trained to output similar embeddings for similar speakers; thus similar speakers tend to end up with the same cluster index $\hat{y}_k$. We can consider the cluster index as a pseudo-speaker label, and have it take the place of the label $y_j$ in the standard ARW adversary formulation.

The adversary network $f_\phi$ is built as a multilayer perceptron network with nonlinear activation. 
The input is the embedding extracted from the anchor utterance, and the output layer has the probabilities of the $K$ classes, $f_\phi(\mathbf{x}^a_j,\hat{y}_j)$ for $\hat{y}_j \in \{0 , \ldots, K-1\}$. 
This leads to
\begin{equation}
J(\theta, \phi) = \min_{\theta} \max_{\phi}  \sum_{\mathbf{x}_{j,k} \in [D]} \left(
  \sum_{j=0}^{N-1}  \lambda^{K}_{\phi} (\mathbf{x}^a_j, \hat{y}_j) L_{p,j} \right) \ , \label{eq:pseudo_labelling}
\end{equation}
where the adversarial weight is computed by normalizing the adversary network outputs:
\begin{equation}
    \lambda^{K}_{\phi} (\mathbf{x}^a_j, \hat{y}_j) =
            1 + \frac{ f_\phi(\mathbf{x}^a_j, \hat{y}_j)}{\frac{1}{N} \sum_{k=0}^{N-1} f_\phi(\mathbf{x}^a_k,\hat{y}_k)} \label{eq:lambda_pseudo_label} \ .
\end{equation}
Unlike in APS, in PL the adversarial network has $K$ output nodes, whereas
$f_\phi(\mathbf{x}^a_j,\hat{y}_j) \in \mathbb{R}$ as in the original ARW. 

\subsubsection{Pairwise weighting (PW)\label{sec:pairwise_weighting}}
\begin{figure}[t]
   \centering
   \includegraphics[width=0.7\linewidth]{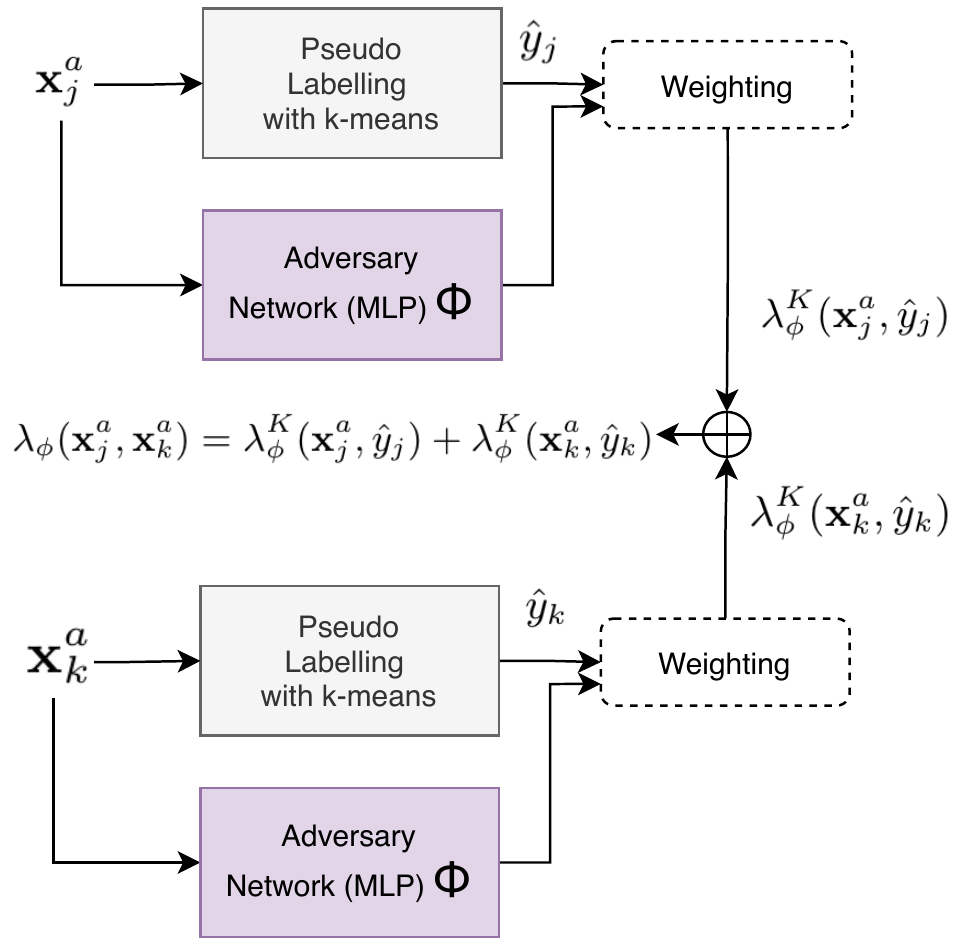}
   \vspace{-.5em}
   \caption{Computing ARW weights $\lambda_{\phi}(\mathbf{x}_j^{a}, \mathbf{x}_k^{a})$ for PW.}
   \label{fig:pw}
\end{figure}
Unlike in the previous two versions of ARW, and as shown in Fig.~\ref{fig:pw}, we formulate two versions of the adversarial weight as a function of speaker {\em pairs}, to match the structure of AP loss:
\begin{multline}
J(\theta, \phi) = \min_{\theta} \max_{\phi}  \\
   \sum_{\mathbf{x}_{j,k} \in [D]} \left( -\frac{1}{N} \sum_{j=0}^{N-1} \mathrm{log} \frac{e^{\lambda_{\phi}(\mathbf{x}_j^{a}, \mathbf{x}_j^{a}) \mathbf{S}_{j,j}}} {\sum_{k=1}^N e^{\lambda_{\phi}(\mathbf{x}_j^{a}, \mathbf{x}_k^{a}) \mathbf{S}_{j,k}}}\right) \ ,
    \label{eq:pairwise_weighting_normalized_weighted}
\end{multline}
and
\begin{multline}
J(\theta, \phi) = \min_{\theta} \max_{\phi}  \\
   \sum_{\mathbf{x}_{j,k} \in [D]} \left( -\frac{1}{N} \sum_{j=0}^{N-1} \mathrm{log} \frac{\lambda_{\phi}(\mathbf{x}_j^{a}, \mathbf{x}_j^{a}) e^{ \mathbf{S}_{j,j}}} {\sum_{k=1}^N \lambda_{\phi}(\mathbf{x}_j^{a}, \mathbf{x}_k^{a}) e^{ \mathbf{S}_{j,k}}}\right) \ .
    \label{eq:pairwise_weighting}
\end{multline}
The first version (\ref{eq:pairwise_weighting_normalized_weighted}) applies the adversarial weight to the similarities between speakers while the second (\ref{eq:pairwise_weighting}) applies it to the exponential of the similarities.
We decompose $\lambda_{\phi}(\mathbf{x}_j^{a}, \mathbf{x}_k^{a})$ as a combination of the single-speaker PL weights (\ref{eq:lambda_pseudo_label}) to avoid doubling the input dimensionality:
\begin{equation}
\lambda_{\phi}(\mathbf{x}_j^{a}, \mathbf{x}_k^{a}) = \lambda^K_{\phi}(\mathbf{x}_j^{a}, \hat{y}_j) + \lambda^K_{\phi}(\mathbf{x}_k^{a}, \hat{y}_k) \ ,
\end{equation}
where $\hat{y}_j$ and $\hat{y}_k$ are the centroid indices computed using (\ref{eq:hat_y_j}).

\section{Experiments\label{sec:experiment}}
\subsection{Experimental setup\label{sec:experimental setup}}
The performance of ARW was evaluated on the VoxCeleb1 database \cite{Nagrani_2017}, with models trained on VoxCeleb2 data \cite{Chung_2018}. 
The training set comprised 5,994 speakers and 1,092,009 utterances. 
For training, we form batches of two utterances each from 200 speakers: $N=200$ and $M=2$.
The backbone model was half ResNet34 from VoxCelebTrainer, with 8M parameters, 4 residual layers and attentive statistical pooling \cite{chung20_odyssey, chung2020defence,  okabe2018attentive}.
The learning rate was 0.001, and for every 3 epochs, the learning rate was decayed by a factor of 0.95.
The evaluation was performed following the protocol from \cite{Chung_2018}.
The speaker embedding size was 512, the default for VoxCelebTrainer.

\subsection{Results\label{sec:experimental_result}}
Table \ref{tab:gender_eer} enumerates the overall and the gender-group EERs of the baseline model, APS, PL, and PW.
For APS, we varied the adversarial network model size from one up to three layers, and layer size (including the output dimension $H$) from 32 to 128 hidden units.
Overall EER ranged from 1.09\% to 1.15\% for APS (\ref{eq:aps_bilinear}) and from 1.12\% to 1.16\% for APS (\ref{eq:aps_exp}); the best result was obtained with one hidden layer of 64 units and 128 units, respectively.
For PL and PW, K-means clustering is performed at the beginning of each epoch with $K=8, 32,$ and $128$.
The adversary network is a 256x3 multilayer perceptron network with sigmoid activation.
In both cases, $K=128$ achieved the lowest EER, namely, 1.08\% for PL and 1.08\% for PW: $K=8$ and $32$ leads to 1.10\% and 1.13\% for PL, and 1.12\% and 1.11\% for PW, respectively.

When comparing APS versions, we observe that the exponential of cosines (\ref{eq:aps_exp}) yielded a smaller gap between groups than inner products (\ref{eq:aps_bilinear}), though the latter had lower overall error.
The EER difference between the baseline and PL is 0.06\% for female and 0.12\% for male speakers.
The improvement for male speakers, which have higher EER, was larger than for female speakers.
The EER gap between male and female speakers decreased from 0.70\% to 0.58\% with PW.
When comparing PW (\ref{eq:pairwise_weighting_normalized_weighted}) and PW (\ref{eq:pairwise_weighting}), we observe that the latter more effectively reduced the EER gap between gender groups.

Table \ref{tab:nat_eer} enumerates the nationality-dependent EERs.
In APS (\ref{eq:aps_bilinear}), we observe that the standard deviation remains similar to the baseline, while APS (\ref{eq:aps_exp}) reduced the standard deviation to 0.19.
We observe that the EER difference between the baseline and PL is 0.02\% for US, 0.03\% for UK, and 0.06\% for other speakers. 
The standard deviation of EERs decreased from 0.21 to 0.20.
Also with PW, the standard deviation is further reduced to 0.19.
The gain for the majority group (US speakers) is modest, and the gain in overall performance is driven by the minority groups (UK and Others). 
The improvement based on nationality breakdown is not as clear as that based on gender.
We conjecture that nationality is a more complex and ambiguous attribute than gender, with a distribution that is highly unbalanced (mostly US and UK).
Also, the Others class lumps together many subgroups with diverse characteristics, and US and UK populations are themselves diverse with substantial overlaps in terms of speech characteristics, making gains harder to detect via our three-way breakdown of the speaker population.




\begin{table}[t!]
\caption{\label{tab:gender_eer} Overall EERs (\%) and gender-group EER (\%) for APS, PL, and PW.}
\centering
\begin{tabular}{l|c|c|c|c}
Method &  ALL & female (45\%) & male  (55\%) & gap \\ \hline
Baseline & 1.17 & 0.69 & 1.39  & 0.70  \\ \hline
APS (\ref{eq:aps_bilinear}) & 1.09 & 0.67  & 1.29   & 0.62  \\ \hline
APS (\ref{eq:aps_exp}) & 1.12 & 0.65 & 1.26 &  0.61  \\ \hline
PL & 1.08& 0.63 & 1.27 & 0.64 \\ \hline
PW (\ref{eq:pairwise_weighting_normalized_weighted}) & 1.09&  0.65 & 1.27 & 0.62 \\ \hline PW(\ref{eq:pairwise_weighting}) & 1.08& 0.67 & 1.25 & 0.58
\end{tabular}
\end{table}


\begin{table}[t!]
\caption{\label{tab:nat_eer}EER by nationality (\%).}
\centering
\begin{tabular}{l|c|c|c|c}
Method & US (64\%) & UK (17\%) & Others (19\%) & std. \\ \hline
Baseline & 1.09 & 0.72 & 1.22 & 0.21  \\ \hline
APS (\ref{eq:aps_bilinear}) & 1.05 & 0.72 & 1.24 & 0.21  \\ \hline
APS (\ref{eq:aps_exp}) & 1.12 & 0.80 & 1.26 & 0.19  \\ \hline
PL & 1.07 & 0.69 & 1.16 & 0.20 \\ \hline
PW (\ref{eq:pairwise_weighting_normalized_weighted}) & 1.09 & 0.80 & 1.26 & 0.19 \\ \hline
PW (\ref{eq:pairwise_weighting}) &1.04 & 0.76 & 1.22 & 0.19
\end{tabular}
\end{table}

\section{Conclusions\label{sec:conclusion}}
We have proposed a novel approach to speaker verification fairness based on adversarial reweighting.
We compared different reweighting algorithms based on adversarial networks, and reduced the EER gap between different groups based on gender and nationality, while achieving an overall EER that is lower than the baseline.
A major advantage of our approach is that while both performance discrepancy and overall error is reduced, no explicit information about group membership is required for model training.
Adversarial reweighting based on pseudo-labelling, as a function of either the anchor speaker alone or of speaker pairs, showed the lowest overall EER.
It is notable that using pairwise weighting can further decrease the standard deviation across different groups.
Future work will include a decoupling of the learner and the adversary model, unlike the current model structure in which the adversary network takes its input from the learner. 
In addition, an analysis on a larger database with balanced nationality groups would be desirable.
Our approach could also be extended to other speech characterization tasks, such as for emotion recognition or medical diagnostics.

\newpage

\bibliographystyle{IEEEtran}

\bibliography{main}

\end{document}